\documentclass[conference, a4paper]{IEEEtran}

\usepackage{cite}
\usepackage{amsmath,amssymb,amsfonts,amsthm,enumitem}
\usepackage[ruled,linesnumbered]{algorithm2e}
\usepackage{algorithmic}
\usepackage{graphicx}
\usepackage{textcomp}
\usepackage{float}
\usepackage{placeins}
\usepackage{multirow}
\usepackage{dblfloatfix}
\usepackage{pifont}
\usepackage[table,xcdraw]{xcolor}
\usepackage[normalem]{ulem}
\useunder{\uline}{\ul}{}
\usepackage{booktabs}
\usepackage[none]{hyphenat}

\usepackage{adjustbox}
\usepackage{enumitem}
\usepackage[english]{babel}
\usepackage[utf8]{inputenc}
\usepackage[colorinlistoftodos]{todonotes}
\usepackage{balance}
\usepackage{comment}

\usepackage{pgfplots}
\usepackage{subcaption}
\begin{document}
	
\title{Security-Aware Approximate Spiking Neural Networks}
\author{\IEEEauthorblockN{Syed Tihaam Ahmad, Ayesha Siddique, Khaza Anuarul Hoque}
\IEEEauthorblockA{\textit{Department of Electrical Engineering and Computer Science}\\ \textit{University of Missouri,
Columbia, MO, USA}\\
tawm9@umsystem.edu, ayesha.siddique@mail.missouri.edu, hoquek@missouri.edu}
}

\maketitle
\begin{abstract}
Deep Neural Networks (DNNs) and Spiking Neural Networks (SNNs) are both known for their susceptibility to adversarial attacks. Therefore, researchers in the recent past have extensively studied the robustness and defense of DNNs and SNNs under adversarial attacks. Compared to accurate SNNs (AccSNN), approximate SNNs (AxSNNs) are known to be up to 4X more energy-efficient for ultra-low power applications. Unfortunately, the robustness of AxSNNs under adversarial attacks is yet unexplored. In this paper, we first extensively analyze the robustness of AxSNNs with different structural parameters and approximation levels under two gradient-based and two neuromorphic attacks. Then, we propose two novel defense methods, i.e., precision scaling and approximate quantization-aware filtering (AQF), for securing AxSNNs. We evaluated the effectiveness of these two defense methods using both static and neuromorphic datasets. Our results demonstrate that AxSNNs are more prone to adversarial attacks than AccSNNs, but precision scaling and AQF significantly improve the robustness of AxSNNs. For instance, a PGD attack on AxSNN results in a 72\% accuracy loss compared to AccSNN without any attack, whereas the same attack on the precision-scaled AxSNN leads to only a 17\% accuracy loss in the static MNIST dataset (4X robustness improvement). Similarly, a Sparse Attack on AxSNN leads to a 77\% accuracy loss when compared to AccSNN without any attack, whereas the same attack on an AxSNN with AQF leads to only a 2\% accuracy loss in the neuromorphic DVS128 Gesture dataset (38X robustness improvement).

\end{abstract}

	

\maketitle 
\vspace{-3mm}
\section{Introduction}
\label{sec:introduction}
Spiking neural networks (SNNs) are the third generation of neural networks that employ event-driven computing capabilities \cite{viale2021carsnn}. In recent years, many SNN models of different sizes have been developed for data analytics, such as gesture recognition, object detection, and image classification. However, large-sized SNN models are known for recognizing more features than small ones. Consequently, the state-of-the-art large-sized SNN models have numerous parameters that need to be considered in both the training and inference phases. This limits the deployment of SNNs on ultra-low power resource-constrained edge devices. To handle this problem, approximate computing in SNNs has recently emerged as an energy-efficient solution. Approximate computing-based SNNs (AxSNNs) relax the abstraction of near-perfect accuracy in error-resilient applications for their low energy consumption. For instance, AxSNNs obtained via approximating the weights can reduce the energy consumption by 4X \cite{sen2017approximate} compared to the accurate SNNs (AccSNNs). 


\begin{figure}[!b]
\centering
\vspace{-0.28in}
\includegraphics[width=0.3\textwidth]{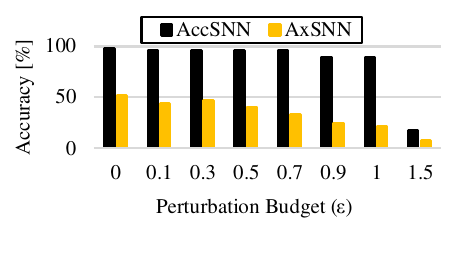}
\vspace{-0.22in}
\caption{Robustness comparison of AccSNN and AxSNN under PGD attack with different perturbation budgets.}
\vspace{-0.02in}
\label{fig:motivational}
\end{figure}

Similar to traditional deep neural networks (DNNs), AxSNNs are also prone to adversarial attacks. Adversarial attacks are known for being very stealthy as they add minimal perturbation noise to the inputs, which are imperceptible to the human eye, yet successfully fool the SNN classifiers. \cite{marchisio2021r}. Motivated by this, the robustness analysis and adversarial defense in AccSNNs have been thoroughly investigated in several recent works \cite{bhattacharjee2022examining,nomura2022robustness,marchisio2021dvs,el2021securing}. Very recently, in \cite{siddique2022approximation}, the authors showed that approximate DNNs (AxDNNs) are more prone to adversarial attacks as compared to accurate DNNs (AccDNN). However, surprisingly, the robustness of AxSNNs is yet unexplored. Therefore, a more comprehensive study is required to understand the inherent behavior of AxSNNs vs. AccSNNs, especially under adversarial attacks. Exploration of such behaviors can enable designing defense techniques tailored specifically for AxSNNs.

\subsection{Motivational Case Study and Key Observations} 
As a preliminary study for motivating our research, we conducted a motivational case study highlighting the impact of adversarial attacks on AcSNNs vs. AccSNNs. For this purpose, we first trained a 5-layered AccSNN, having 3 convolutional layers and 2 fully-connected layers for classifying the MNIST \cite{cohen2017emnist} dataset. Then, we built an AxSNN (using approximation level 0.1) as an approximate counterpart of the AccSNN. Finally, we compared the performance of AccSNN and AxSNN under the $l_\infty$ norm-based projected gradient descent (PGD) attack by varying the perturbation budget $\epsilon$ ranging from 0 to 1.0. The results are presented in Fig. \ref{fig:motivational}. We observe that the AxSNN is significantly less robust when compared to AccSNNs under attack. For instance, when there is no attack ($\epsilon$=0), the accuracy of AccSNN and AxSNN is 97\% and 52\%, respectively. However, for ($\epsilon$=0.5), the accuracy of AccSNN and AxSNN is 95\% and 40\%, respectively, which shows a 55\% difference in their accuracy. Furthermore, when the perturbation budget was varied to 1.0 ($\epsilon$=1.0), we can observe a 68\% difference while comparing the accuracy of the AccSNN and AxSNN. These outcomes motivated us to thoroughly investigate the robustness of the AxSNNs and explore potential defense techniques to design robust AxSNNs.



\subsection{Novel Contributions} In this paper, we present a security-aware AxSNNs design method with the following novel contributions:
\setenumerate{leftmargin =*}
\begin{enumerate}
    
    \item A novel design approach for designing adversarially robust AxSNNs by identifying their robustness-aware knobs through \textit{precision scaling}, i.e., through finding the appropriate combination approximation levels, structural parameters (threshold voltage, time steps), and quantization. We also propose a defense method through Approximate Quantization-aware filtering (AQF), specifically effective against neuromorphic attacks.  \textbf{[Section \ref{sec:methodology}]}
    
    \item An extensive vulnerability analysis of AxSNNs vs. AccSNNs against two gradient-based attacks and two neuromorphic attacks under different threshold voltage, time steps, precision scales, and approximation levels. Specifically, we evaluated the impact of Projected Gradient Descent (PGD) and Basic Iterative Method (BIM) attacks on AccSNN and AxSNN with a static dataset MNIST \cite{cohen2017emnist}. On the contrary, we evaluated the impact of Sparse and Frame attacks on AccSNN and AxSNN with the neuromorphic dataset DVS128 Gesture \cite{amir2017low}). \textbf{[Section \ref{sec:results}]}
\end{enumerate}


Our results demonstrate that AxSNNs are more prone to adversarial attacks than AccSNNs. However, the precision scaling and approximate quantization aware filtering improve their robustness significantly. For instance, a PGD attack with perturbation budget 1.0 on AxSNNs results in a 72\% accuracy loss, whereas the same attack on AccSNNs results in only a 9\% accuracy loss in MNIST classification. Interestingly, the same AxSNN with precision scaling shows only a 17\% accuracy loss, indicating a 4X robustness improvement. Similarly, a Sparse Attack on AxSNN leads to a 77\% accuracy loss when compared to AccSNN without any attack in the DVS128 Gesture classification. However, after using our designed approximate quantization aware filter the accuracy loss is just 2\%, indicating a 38X improvement in robustness.

\section{Preliminaries}
\label{sec:prelim}
This section provides a brief overview of SNNs and adversarial attacks to understand the paper better.

\textbf{Approximate Spiking Neural Networks:} 
AxSNNs employ approximate computing to trade their classification accuracy with energy efficiency in ultra-low power applications. AxSNNs typically associate an approximation level $a_{th}$ with each spiking neuron. The $a_{th}$ determines if the respective neurons should be activated or deactivated based on the sensitivity of the neurons to errors and spiking activity \cite{venkataramani2014axnn}. Similar to AccSNNs, AxSNNs use the standard leaky-integrate-and-fire (LIF) model where, when the membrane potential exceeds the threshold voltage, the neuron emits an output spike and resets its membrane potential. Specifically, they process the spike-encoded inputs which are most commonly encoded using rate encoding, where the activation activity corresponds to the mean firing rates of spikes over certain time steps. The time steps refer to the observation period when the SNN receives the same input. 

\textbf{Adversarial Attacks:} The adversarial attacks are small perturbations that cause the classifier to predict the false labels. Examples of such attacks include gradient-based PGD and BIM, which are strong attacks in the adversarial machine learning domain. The gradient-based attacks exploit the concept of back-propagation; however, instead of calculating the gradient with respect to the weights of the model, they craft the adversarial examples by perturbing the input images. Recent studies show that these attacks cannot be used for perturbing the neuromorphic datasets due to their event-driven nature. Therefore, specialized neuromorphic attacks, such as Sparse and Frame Attacks are used \cite{marchisio2021dvs}. A Sparse Attack is a stealthy attack that iteratively perturbs the neuromorphic images based on the loss function of output label probability to generate perturbed events. On the other hand, a Frame Attack is a simple yet effective neuromorphic attack that generates perturbed events by attacking every pixel of the boundary for all the events. 

\vspace{-2mm}
\section{Threat Model}
\vspace{-2mm}
\label{sec:threatModel}
In this section, we present a threat model for exploring the adversarial robustness of AxSNNs.
\label{subsec:advknow}

\textbf{Adversary's Knowledge:} We assume that the adversary uses an accurate classifier model for crafting the adversarial examples. Furthermore, the adversary has partial knowledge about AxSNN, i.e., the internal architecture of the classifier model is known, but the inexactness and model parameters such as threshold voltage and time-steps, precision scale, and approximation levels are not known to the adversary.

\textbf{Attack Generation:} The adversary is assumed capable of evading the classifier model by tampering the input images in the prediction phase without influencing the training data. The adversary crafts the adversarial examples by finding the perturbations that maximize the loss of a model on input while keeping the perturbation magnitude lower than the perturbation budget $\epsilon$. As mentioned earlier, we use iterative gradient-based attacks specifically, $l_\infty$ norm-based \textit{BIM and PGD}, which are considered high-strength attacks for the static datasets. 
We also employ neuromorphic attacks specifically, sparse and frame attacks which are stealthy yet effective in perturbing the neuromorphic images with high resolutions. We used $\epsilon$=1.0 in this paper because the accuracy of both AccSNN and AxSNN drops significantly after this value, so it becomes non-recoverable. As an example, it drops to 10\% with $\epsilon$=1.5.

\section{Security-Aware AxSNN Design Approach}
\label{sec:methodology}

In this section, we discuss our proposed approach for designing a robust and secure AxSNN in detail.

\subsection{Precision-scaling} 
Traditionally, the approximation levels are determined by identifying the maximum tolerable perturbation (MTP) in a neural network \cite{khalid2019qusecnets}. However, this becomes challenging in precision-scaled AxSNNs due to the variation in accuracy with the change in the threshold voltage, time steps and precision scales. Intuitively, an increase in the number of time steps and threshold voltage may lead to a higher number of insignificant spikes by some neurons and hence, affect the classification accuracy. Skipping such neurons has the potential to improve the accuracy of AxSNNs; however, their robustness can decrease under attacks. 
Since precision scaling has the potential to improve the robustness of AccDNNs \cite{khalid2019qusecnets}, determining approximation on the basis of the precision scales in AxSNNs can improve the adversarial robustness. However, exploring a precision scale in addition to threshold voltage and time steps for robust approximation seems challenging. In this paper, we determine the  approximation levels $a_{th}$ in AxSNNs by using the following equation: 
\vspace{-3mm}
\begin{equation}
\label{eq:precision}
a_{th} = (cN_s/T) \cdot min(1,  V_m / V_{th}) \cdot \sum_{i=1}^{c} w_i^{p},
\vspace{-2mm}
\end{equation}

\noindent where $c$, $N_s$, $T$, $V_m$, $V_{th}$ and $w^{p}$ denote the number of connections to output, number of spikes, time steps, membrane potential, threshold voltage, and precision scaled weight of neuron. Furthermore, $min (1, V_m/ V_{th})$ is the spike probability. The maximum spike probability is 1 when $V_m$ crosses $V_{th}$ otherwise, $V_m/ V_{th}$. The spike probability is weighted by the mean of all weights corresponding to a connection $c$ i.e., $\sum_{i=1}^{c} w_i^{p}$ which includes the precision scaling of weights. 

Algorithm 1 delineates the steps involved in implementing this equation for robustness in AxSNNs. We first initialize a counter $adv$ for successful attack generation (Line 1). Then, we train an AccSNN model with the given threshold voltage $v$ and time steps $t$ and save the trained model and the corresponding weights of each layer l (Line 3). This learning phase is quantitatively verified with a quality constraint $Q$ i.e., minimum baseline accuracy below which we consider SNN learning inefficient (Line 4). The value of $Q$ depends on the given SNN architecture, dataset, and application. Then, we craft the adversarial examples for an adversarial attack and perturbation budget $\epsilon$ (Line 5). The adversarial defense through precision scaling starts from Line 6 where the trained weights are sorted initially in ascending order. This sorting helps us in approximating the weights according to their significance later on. Afterwards, we perform precision scaling on the basis of each precision scale $s$ and calculate the mean of all connections in layer $l$ (Lines 8-9). Using this mean, we determine $a_{th}$ (as discussed earlier) for each layer $l$ and approximate the precision-scaled model by removing the connections having weights below $a_{th}$ (Line 10). If a neuromorphic dataset is given as an input and the corresponding flag $F_d$ is high then, we also use a special approximate quantization aware mask filter (AQF) from Algorithm 2 as discussed in Section \ref{subsec:aqaf} (Line 12-14). Next, the algorithm checks if the crafted adversarial example can fool AxSNN, i.e., if the attack succeeds in forcing the output to a wrong label, and accordingly increment the adversarial success counter (Line 15-18). Lastly, we evaluate the robustness for the perturbation budget $\epsilon$ as the rate of attacks for which the adversary failed to generate an adversarial example that fools the victim SNN (Line 21). In this paper, we use this algorithm to find the approximation level $a_{th}$ with the robust set of threshold voltage, time steps, and precision scales. Therefore, we compare the accuracies of the precision-scaled AxSNN models across all these parameters and return their values which meet our quality constraint (Lines 22-24).\\

\SetInd{0.5em}{0.5em}
\begin{algorithm}[!ht]
	\caption{Precision-Scaling in AxSNNs}
	\label{alg:robust}
        \small
		\footnotesize
        \DontPrintSemicolon

        \SetKwInOut{Input}{Inputs}\SetKwInOut{Output}{Outputs}
        \Input{Type of adversarial attack: $attack$; \\
        Perturbation budget: $\epsilon$; \\
        Time steps: $T$ = $[t_1, t_2, ..., t_n]$; \\
        Threshold Voltage: $V_{th}$ = $[v_1, v_2, ... v_n];$ \\
        Train dataset: $\mathcal{D}_{tr}$ = ($X$, $L$); \\
        Test dataset: $\mathcal{D}_{ts}$ = ($x$, $l$); \\
        Perturbation budget: $\epsilon$; \\
        Precision-scaling level: $s_{l} = [s_1, s_2, ...,s_n]$ ;\\
        Quality constraint: $Q$; \\
        Neuromorphic Dataset Flag: $F_d$
        }

		\Output{Robustness level $R$, Best $V_{th}$, time steps $ts$, approximation level $a_{th}$ and precision-scaling level $s$}
 
		\begin{algorithmic}[1]
            \STATE $adv$ = 0;
            \FOR {\textbf{each} ($v$, $ts$) in ($V_{th}$, $T$)}{
                \STATE ($model$, $w_l$) = trainAccurateSNN ($v$, $ts$, $\mathcal{D}_{tr}$)\\
                \IF {Accuracy ($model$) $\ge$ $Q$}{
                    \STATE ($x^{*}_k$, $l^{*}_k$) = AdvExGen ($model$, $\epsilon$, $attack$, $X^t_k$) \\ 
                    \STATE $w$ = SortInAscendingOrder ($w_l$); \\
                    \FOR {\textbf{each} $s$ in $s_l$}{
                        \STATE $w^p$ = PrecisionScaling ($w$, s); \\
                        \STATE $m_l^c$ = $\sum_{j=1}^{c} (w_j^p)$; \\
                        \STATE $a_{th}$ = ($cN_s/T$) $m_l^c$ $\cdot$ min (1, $V_m$ / $v$) \\
                        \STATE $model$ = ApproximateSNN($model$, $w^p$, $m_l^c$ $a_{th}$); \\
                        \IF{$T_d$ is TRUE}{
                            \STATE $\mathcal{D}_{ts}$ = ApproximateQuantizedFilter($q_t$, $\mathcal{D}_{ts}$)\\
                        }\ENDIF
                        \STATE ($x'_k$, $l'_k$) = AdvAttacks ($model$, $\epsilon$, $attack$, $x^{*}_k$, $l^{*}_k$) \\ 
                        \IF {$l'_k$ $\neq$ $l_k$}{
                            \STATE adv++;}
                        \ELSE{
                            \STATE NOP;
                        }\ENDIF
                        \STATE $R$ ($\epsilon_i$) = (1 - $adv$/ size($\mathcal{D}_{ts}$)) * 100; \\
                        \IF {$R$ $\ge$ $Q$}{
                        \RETURN ($R$, $V_{th}$, $T$, $s$, $a_{th}$)
                        }\ENDIF
                    }\ENDFOR
                }\ENDIF
        }\ENDFOR
      
\end{algorithmic}
\end{algorithm}

\subsection{Approximate Quantization Aware Filtering}
\label{subsec:aqaf}
\vspace{-1mm}
For SNNs feeded by dynamic vision sensors (DVS), the above-discussed defense techniques for frame-based sensors cannot be directly applied due to the event-driven nature of neuromorphic images. Therefore, we present an additional approximate quantization aware filter (AQF) to remove uncorrelated events from the neuromorphic images. Our proposed Algorithm \ref{alg:aqaf} removes adversarial perturbation noise from event-driven neuromorphic dataset $E$. The neuromorphic dataset $E$ is represented in the form of $(x, y, p, t)$, where $x$, $y$, $p$ and $t$ denote the x-coordinate, the y-coordinate, the polarity and the timestamp of the event $E$ respectively. The events $e$ are associated with a spatio-temporal domain and hence, they are correlated. Our algorithm calculates the correlation between events $e$. If the correlation for events $e$ is lower than certain spatial-temporal thresholds ($s$, $T1$, $T2$) then, they are removed because these events with very low correlation are more likely to be noisy due to adversarial perturbations. 

\SetInd{0.5em}{0.5em}
\begin{algorithm}[!h] \small
	\caption{ApproximateQuantizedFilter}
	\label{alg:aqaf}
        \algsetup{linenosize=\small}
		\footnotesize
        \DontPrintSemicolon

        \SetKwInOut{Input}{Inputs}\SetKwInOut{Output}{Outputs}
        \Input{List of events: $D_{ts}$$(x,y,p,t) = Events$\\
        
        Quantization step: $qt$;\\
        }
		\Output{Filtered Quantized dataset $D_q$}
		\begin{algorithmic}[1]
        \STATE $M = 0$
        \STATE $activity=0$, $s=2$, $T1=5$ , $T2=50$
        \FOR {$e$ in $Events$ }{
        \STATE $e$ = round($e$/$q_t$) $\cdot$ $q_t$
            \FOR {$i$ in $(x_e-s,x_e+s)$)}{
                 \FOR {$j$ in $(y_e-s,y_e+s)$ )}{
                    \IF{not($i==x_e$ and $j==y_e$)}{
                        \STATE $M[i][j] = t_e$
                    }\ENDIF
                    \IF{not($i==x_e$ and $j==y_e$)}{
                        \STATE $activity[i][j] += 1$
                    }\ENDIF
                }\ENDFOR
            }\ENDFOR
            \IF{$activity[i][j] > T1$}{
                        \STATE $M[i][j] = 1$
                }\ENDIF
            \IF{$t_e-M[x_e][y_e]>T2$ or $M[x_e][y_e]==1$)}{
               \STATE Remove $e$ from $Events$
            }\ENDIF
        }\ENDFOR 
        \STATE $D_q$ = $Events$
        \RETURN $D_q$
        
\end{algorithmic}
\end{algorithm}

Algorithm \ref{alg:aqaf} delineates the steps involved in approximate quantization aware filtering of perturbed events in the neuromorphic dataset. First, the dataset is quantized with a fixed quantization step $q_t$ (Line 4) then, the uncorrelated values are checked in each event $e$ of the dataset (Lines 5-9). With each low correlation, a counter variable $activity$ is increased and flagged (Lines 10-16). Finally, the flagged uncorrelated values events $e$ are removed from the dataset (Lines 18-20) to get a quantized and filtered dataset $D_q$ (Line 23).
\vspace{-0.04in}



\vspace{-1mm}
\section{Results and Discussions}
\label{sec:results}
\vspace{-2mm}
In this section, we present our results for the adversarial vulnerability analysis and adversarial defense in AxSNNs. 

\vspace{-2mm}
\subsection{Datasets and Architectures} 
\vspace{-2mm}
We use both static and neuromorphic datasets i.e., MNIST \cite{cohen2017emnist} and DVS128 Gesture \cite{amir2017low}. Both of these datasets are common for evaluating the performance of SNNs~\cite{Putra2021QSpiNNAF} in embedded platforms. As a classifier for the MNIST dataset, we employed a 7-layered SNN with three convolutional layers, two pooling, and two fully connected layers. The test accuracy of this architecture on clean inputs (without any adversarial attack) is 97\%. On the other hand, as a classifier for the DVS128 Gesture dataset, we employed an 8-layered SNN with two convolutional and two fully connected layers, three pooling, and one dropout layer. The test accuracy of this architecture on clean inputs is 92\%.

\begin{figure}[!h]
	\centering
	\includegraphics[width=0.8\linewidth]{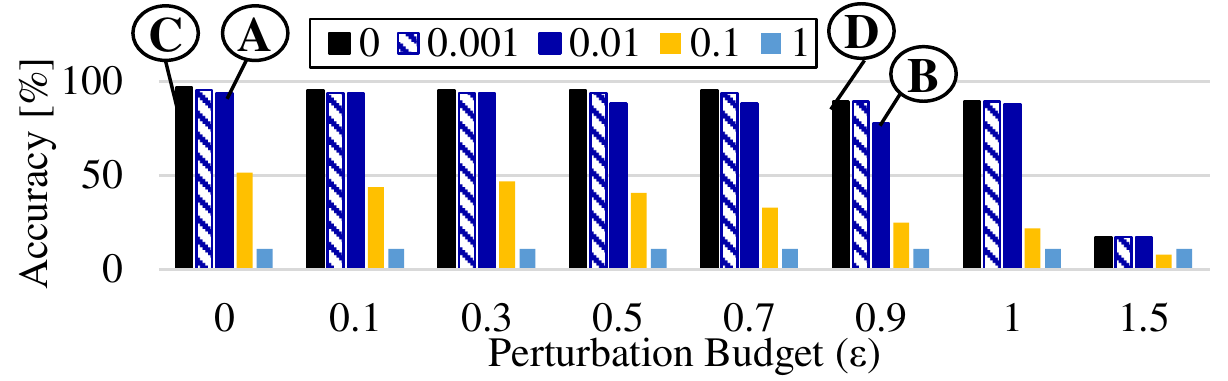}
	\vspace{-1mm}
    \caption{Robustness analysis of AccSNN (approximation level 0) and AxSNN MNIST classifier under PGD attack for approximation levels 0.001, 0.01, 0.1 and 1.}
	\label{fig:approxlevelpgd}
	\vspace{-0.1in}
\end{figure}

\begin{figure}[!h]
	\centering
	\includegraphics[width=0.85\linewidth]{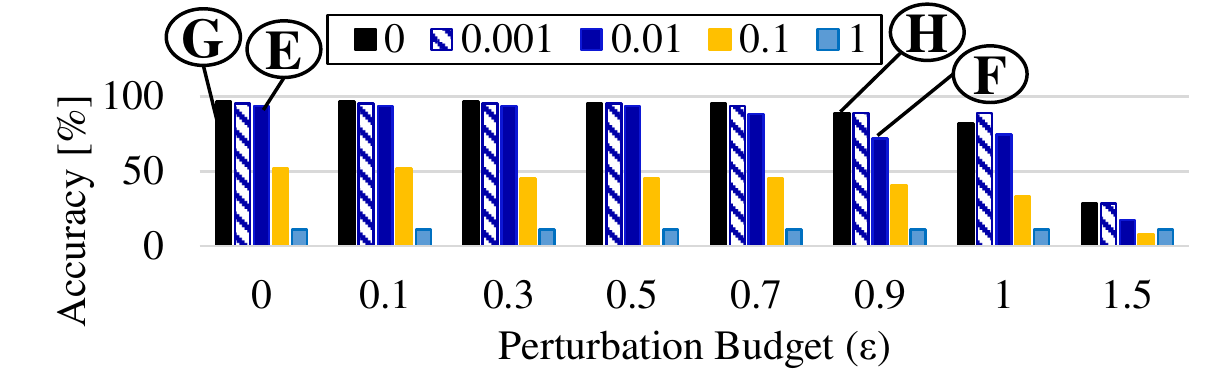}
	\vspace{-2mm}
    \caption{Robustness analysis of AccSNN (approximation level 0) and AxSNN MNIST classifier under BIM attack for approximation levels 0.001, 0.01, 0.1 and 1.}
	\label{fig:approxlevelbim}
	\vspace{-0.1in}
\end{figure}

\subsubsection{\uline{Vulnerability analysis of AxSNNs}}
We first evaluate the adversarial robustness of AxSNNs by varying the approximation levels and comparing them with AccSNNs under different perturbation budgets for the static MNIST dataset. We keep threshold voltage and time steps constant at 0.25 and 32 in this experiment. Our results in Fig. \ref{fig:approxlevelpgd} and Fig. \ref{fig:approxlevelbim} show that the robustness of AxSNNs decreases with an increase in the perturbation budget. For example, a PGD attack ($\epsilon$ = 0.9) drops the accuracy of AxSNN with approximation level 0.01 from 93\% ($\epsilon$ = 0) to 77\%  (see labels A and B in Fig. \ref{fig:approxlevelpgd}) whereas the same attack drops the accuracy of the AccSNN from 96\% to 89\% see Label C and D in Fig. \ref{fig:approxlevelpgd}. This indicates a 7\% accuracy loss for the AccSNN under PGD attack, whereas the same attack causes accuracy loss of 16\% for the AxSNN. In this case, the AxSNN is 2X more vulnerable than AccSNN. Likewise, a BIM attack ($\epsilon$ = 0.9) drops the accuracy of the AxSNN with approximation level 0.01 from 93\% ($\epsilon$ = 0) to 71\%  (see labels E and F in Fig. \ref{fig:approxlevelbim}) whereas the same attack drops the accuracy of the AccSNN from 96\% to 82\% (see Label G and H in Fig. \ref{fig:approxlevelbim}). This indicates a 14\% accuracy drop for the AccSNN, whereas the accuracy drop for the AxSNN is around 22\% with PGD attack. In this case, the AxSNN is 1.5X more vulnerable than AccSNN.

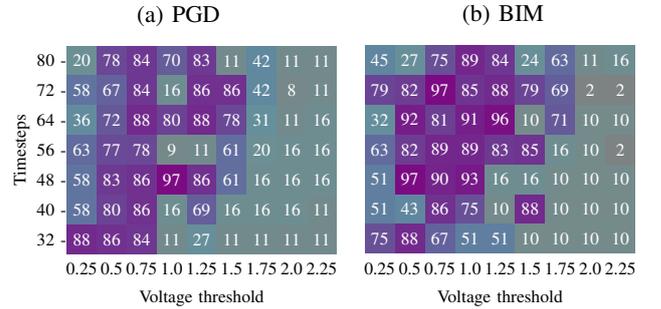
\begin{figure}[!t]
\centering
	\begin{subfigure}[b]{.2475\textwidth}
		\centering
        \caption{PGD}
		{\resizebox{\textwidth}{!}{\definecolor{mycolor2}{rgb}{0.00000,0.44700,0.7499}%
\definecolor{mycolor1}{rgb}{0.85000,0.32500,0.09800}%
\definecolor{mycolor3}{rgb}{0,128,0}%

\begin{tikzpicture}[scale=0.6]

\node at  (-1,-4) [rotate=90] {Timesteps};

\node at (0,-1) {80 -};
\node at (0,-2) {72 -};
\node at (0,-3) {64 -};
\node at (0,-4) {56 -};
\node at (0,-5) {48 -};
\node at (0,-6) {40 -};
\node at (0,-7) {32 -};
\node at (1,-8) {0.25};
\node at (2,-8) {0.5};
\node at (3,-8) {0.75};
\node at (4,-8) {1.0};
\node at (5,-8) {1.25};
\node at (6,-8) {1.5};
\node at (7,-8) {1.75};
\node at (8,-8) {2.0};
\node at (9,-8) {2.25};
\node at (5,-9) {Voltage threshold};

\foreach \y [count=\n] in {
      {20,78,84,70,83,11,42,11,11},
      {58,67,84,16,86,86,42,8,11},
      {36,72,88,80,88,78,31,11,16},
      {63,77,78,9,11,61,20,16,16},
      {58,83,86,97,86,61,16,16,16},
      {58,80,86,16,69,16,16,16,11},
      {88,86,84,11,27,11,11,11,11}, 
    } {
      \foreach \x [count=\m] in \y {
        \node[fill=violet!\x!violet!\x!violet!\x!cyan!\x!gray, minimum size=6mm, text=white] at (\m,-\n) {\x};  
      }
    }
\end{tikzpicture}}\label{fig:pgdfp32}}
	\end{subfigure}
     \begin{subfigure}[b]{.21\textwidth}
    		\centering
            \caption{BIM}
    		{\resizebox{\textwidth}{!}{\definecolor{mycolor2}{rgb}{0.00000,0.44700,0.7499}%
\definecolor{mycolor1}{rgb}{0.85000,0.32500,0.09800}%
\definecolor{mycolor3}{rgb}{0,128,0}%

\begin{tikzpicture}[scale=0.6]
\node at (1,-8) {0.25};
\node at (2,-8) {0.5};
\node at (3,-8) {0.75};
\node at (4,-8) {1.0};
\node at (5,-8) {1.25};
\node at (6,-8) {1.5};
\node at (7,-8) {1.75};
\node at (8,-8) {2.0};
\node at (9,-8) {2.25};
\node at (5,-9) {Voltage threshold};

\foreach \y [count=\n] in {
      {45,27,75,89,84,24,63,11,16},
      {79,82,97,85,88,79,69,2,2},
      {32,92,81,91,96,10,71,10,10},
      {63,82,89,89,83,85,16,10,2},
      {51,97,90,93,16,16,10,10,10},
      {51,43,86,75,10,88,10,10,10},
      {75,88,67,51,51,10,10,10,10}, 
    } {
      \foreach \x [count=\m] in \y {
        \node[fill=violet!\x!violet!\x!violet!\x!cyan!\x!gray, minimum size=6mm, text=white] at (\m,-\n) {\x};  
      }
    }
\end{tikzpicture}}\label{fig:bimfp32}}
    	\end{subfigure}
    \vspace{-0.1in}
    \caption{Accuracy of AxSNN (approx. level 0.01, precision-scale FP32) under attack ($\epsilon$ = 1) for MNIST.} 
    \label{fig:fp32}
    \vspace{-0.15in}

\end{figure}

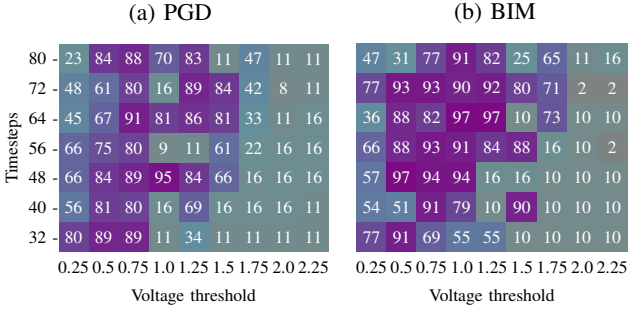
\begin{figure}[!t]
\centering
     \begin{subfigure}[b]{.2475\textwidth}
    		\centering
            \caption{PGD}
    		{\resizebox{\textwidth}{!}{\definecolor{mycolor2}{rgb}{0.00000,0.44700,0.7499}%
\definecolor{mycolor1}{rgb}{0.85000,0.32500,0.09800}%
\definecolor{mycolor3}{rgb}{0,128,0}%

\begin{tikzpicture}[scale=0.6]


\node at (-1,-4) [rotate=90]{Timesteps};
\node at (0,-1) {80 -};
\node at (0,-2) {72 -};
\node at (0,-3) {64 -};
\node at (0,-4) {56 -};
\node at (0,-5) {48 -};
\node at (0,-6) {40 -};
\node at (0,-7) {32 -};
\node at (1,-8) {0.25};
\node at (2,-8) {0.5};
\node at (3,-8) {0.75};
\node at (4,-8) {1.0};
\node at (5,-8) {1.25};
\node at (6,-8) {1.5};
\node at (7,-8) {1.75};
\node at (8,-8) {2.0};
\node at (9,-8) {2.25};
\node at (5,-9) {Voltage threshold};
\foreach \y [count=\n] in {
      {23,84,88,70,83,11,47,11,11},
      {48,61,80,16,89,84,42,8,11},
      {45,67,91,81,86,81,33,11,16},
      {66,75,80,9,11,61,22,16,16},
      {66,84,89,95,84,66,16,16,16},
      {56,81,80,16,69,16,16,16,11},
      {80,89,89,11,34,11,11,11,11}, 
    } {
      \foreach \x [count=\m] in \y {
        \node[fill=violet!\x!violet!\x!violet!\x!cyan!\x!gray, minimum size=6mm, text=white] at (\m,-\n) {\x}; 
      }
    }
\end{tikzpicture}}\label{fig:pgdfp16}}
    	\end{subfigure}
     \begin{subfigure}[b]{.21\textwidth}
		\centering
        \caption{BIM}
		{\resizebox{\textwidth}{!}{\definecolor{mycolor2}{rgb}{0.00000,0.44700,0.7499}%
\definecolor{mycolor1}{rgb}{0.85000,0.32500,0.09800}%
\definecolor{mycolor3}{rgb}{0,128,0}%

\begin{tikzpicture}[scale=0.6]
\node at (1,-8) {0.25};
\node at (2,-8) {0.5};
\node at (3,-8) {0.75};
\node at (4,-8) {1.0};
\node at (5,-8) {1.25};
\node at (6,-8) {1.5};
\node at (7,-8) {1.75};
\node at (8,-8) {2.0};
\node at (9,-8) {2.25};
\node at (5,-9) {Voltage threshold};

\foreach \y [count=\n] in {
      {47,31,77,91,82,25,65,11,16},
      {77,93,93,90,92,80,71,2,2},
      {36,88,82,97,97,10,73,10,10},
      {66,88,93,91,84,88,16,10,2},
      {57,97,94,94,16,16,10,10,10},
      {54,51,91,79,10,90,10,10,10},
      {77,91,69,55,55,10,10,10,10}, 
    } {
      \foreach \x [count=\m] in \y {
        \node[fill=violet!\x!violet!\x!violet!\x!cyan!\x!gray, minimum size=6mm, text=white] at (\m,-\n) {\x};  
      }
    }
\end{tikzpicture}}\label{fig:bimfp16}}
	\end{subfigure}
    \vspace{-2mm}
    \caption{Accuracy of AxSNN (approx. level 0.01, precision-scale FP16 under attack ($\epsilon$ = 1) for MNIST.} 
    \label{fig:fp16}
    \vspace{-0.3in}
\end{figure}
Furthermore, we observe that the robustness of AxSNNs varies with the change in the approximation levels. For instance, under no attack ($\epsilon$=0), the accuracy of AxSNNs drops from 96\% to 93\%, 51\%, and 10\% when we change the approximate level from 0 to 0.01, 0.1 and 1.0, respectively. Conversely, the accuracy of AxSNNs drops from 96\% ($\epsilon$=0) to 77\%, 25\%, and 10\% when we change the approximation levels from 0 to 0.01, 0.1 and 1.0, respectively under a PGD attack ($\epsilon$=0.9) (see Fig. \ref{fig:approxlevelpgd}). A similar trend is also observed for the BIM attack (see Fig. \ref{fig:approxlevelbim}). 

Next, we explore the vulnerability of AxSNNs under sparse and frame attacks with the neuromorphic DVS128 Gesture dataset. We keep the threshold voltage and time steps constant to 1.0 and 80 in this experiment. Our results demonstrate that AxSNNs are vulnerable to neuromorphic attacks. For instance, the accuracy of AxSNN drops from 92\% to 12\% and 10\% under sparse and frame attacks which is almost the same case in AccSNNs as shown in Fig. \ref{fig:noattackandgraph}.




\subsubsection{\uline{Precision-scaling-based Adversarial Defense}}
\begin{figure}[!t]
\centering
     \begin{subfigure}[b]{.2475\textwidth}
		\centering
        \caption{PGD}
		{\resizebox{\textwidth}{!}{\definecolor{mycolor2}{rgb}{0.00000,0.44700,0.7499}%
\definecolor{mycolor1}{rgb}{0.85000,0.32500,0.09800}%
\definecolor{mycolor3}{rgb}{0,128,0}%

\begin{tikzpicture}[scale=0.6]
\node at (0,-1) {80 -};
\node at (0,-2) {72 -};
\node at (0,-3) {64 -};
\node at (0,-4) {56 -};
\node at (0,-5) {48 -};
\node at (0,-6) {40 -};
\node at (0,-7) {32 -};
\node at (-1,-4) [rotate=90]{Timesteps};
\node at (1,-8) {0.25};
\node at (2,-8) {0.5};
\node at (3,-8) {0.75};
\node at (4,-8) {1.0};
\node at (5,-8) {1.25};
\node at (6,-8) {1.5};
\node at (7,-8) {1.75};
\node at (8,-8) {2.0};
\node at (9,-8) {2.25};
\node at (5,-9) {Voltage threshold};
\foreach \y [count=\n] in {
      {22,81,86,70,83,11,47,11,11},
      {47,53,83,16,89,84,42,8,11},
      {31,70,91,81,86,81,33,11,16},
      {66,78,73,9,11,61,22,16,16},
      {64,83,86,95,84,66,16,16,16}, 
      {56,80,81,16,69,16,16,16,11},
      {80,91,92,11,34,11,11,11,11}, 
    } {
      \foreach \x [count=\m] in \y {
        \node[fill=violet!\x!violet!\x!violet!\x!cyan!\x!gray, minimum size=6mm, text=white] at (\m,-\n) {\x};
        }
    }
\end{tikzpicture}}\label{fig:pgdint8}}
	\end{subfigure}
	\begin{subfigure}[b]{.21\textwidth}
		\centering
        \caption{BIM}
		{\resizebox{\textwidth}{!}{\definecolor{mycolor2}{rgb}{0.00000,0.44700,0.7499}%
\definecolor{mycolor1}{rgb}{0.85000,0.32500,0.09800}%
\definecolor{mycolor3}{rgb}{0,128,0}%

\begin{tikzpicture}[scale=0.6]

\node at (1,-8) {0.25};
\node at (2,-8) {0.5};
\node at (3,-8) {0.75};
\node at (4,-8) {1.0};
\node at (5,-8) {1.25};
\node at (6,-8) {1.5};
\node at (7,-8) {1.75};
\node at (8,-8) {2.0};
\node at (9,-8) {2.25};
\node at (5,-9) {Voltage threshold};

\foreach \y [count=\n] in {
      {51,32,76,91,83,25,65,11,16},
      {78,94,94,91,92,80,71,2,2},
      {36,89,84,97,97,10,73,10,10},
      {66,88,93,90,85,88,16,10,2},
      {56,97,94,96,16,16,10,10,10},
      {54,51,91,79,10,90,10,10,10},
      {79,90,69,53,53,10,10,10,10}, 
    } {
      \foreach \x [count=\m] in \y {
        \node[fill=violet!\x!violet!\x!violet!\x!cyan!\x!gray, minimum size=6mm, text=white] at (\m,-\n) {\x};  
      }
    }
\end{tikzpicture}}\label{fig:bimfp16}}
	\end{subfigure}
    \vspace{-2mm}
    \caption{Accuracy of AxSNN (approx. level 0.01, precision-scale INT8) under attack ($\epsilon$ = 1) for MNIST.} 
    \label{fig:int8}
    \vspace{-0.2in}
\end{figure}
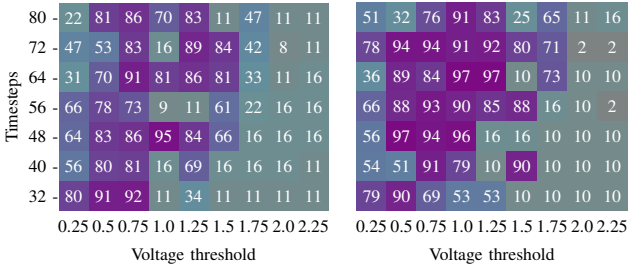

\begin{figure}[!t]
\centering
	\begin{subfigure}[b]{.2175\textwidth}
		\vspace{-0.1in}
          \caption{MNIST}
		\vspace{0.06in}
		{\resizebox{\textwidth}{!}{\definecolor{mycolor2}{rgb}{0.00000,0.44700,0.7499}%
\definecolor{mycolor1}{rgb}{0.85000,0.32500,0.09800}%
\definecolor{mycolor3}{rgb}{0,128,0}%

\begin{tikzpicture}[scale=0.6]

\node at (5,-9) {Voltage threshold};
\node at (-1,-4) [rotate=90]{Timesteps};

\node at (0,-1) {80 -};
\node at (0,-2) {72 -};
\node at (0,-3) {64 -};
\node at (0,-4) {56 -};
\node at (0,-5) {48 -};
\node at (0,-6) {40 -};
\node at (0,-7) {32 -};
\node at (1,-8) {0.25};
\node at (2,-8) {0.5};
\node at (3,-8) {0.75};
\node at (4,-8) {1.0};
\node at (5,-8) {1.25};
\node at (6,-8) {1.5};
\node at (7,-8) {1.75};
\node at (8,-8) {2.0};
\node at (9,-8) {2.25};

\foreach \y [count=\n] in {
      {99,99,94,94,97,24,88,97,16},
      {93,97,99,91,83,97,97,99,11},
      {94,99,97,91,96,10,97,10,10},
      {99,99,99,99,93,96,97,10,11},
      {99,99,99,10,16,16,10,10,10},
      {97,71,97,94,10,97,47,10,10},
      {97,10,96,86,66,10,16,10,10}, 
    } {
      \foreach \x [count=\m] in \y {
        \node[fill=violet!\x!violet!\x!violet!\x!cyan!\x!gray, minimum size=6mm, text=white] at (\m,-\n) {\x};  
      }
    }
\end{tikzpicture}}
		\label{fig:noattack}}
	\end{subfigure}
	\begin{subfigure}[b]{.2625\textwidth}
		\vspace{-0.1in}		
        \caption{DVS128Gesture}
        \vspace{0.05in}
		\includegraphics[width=1\textwidth]{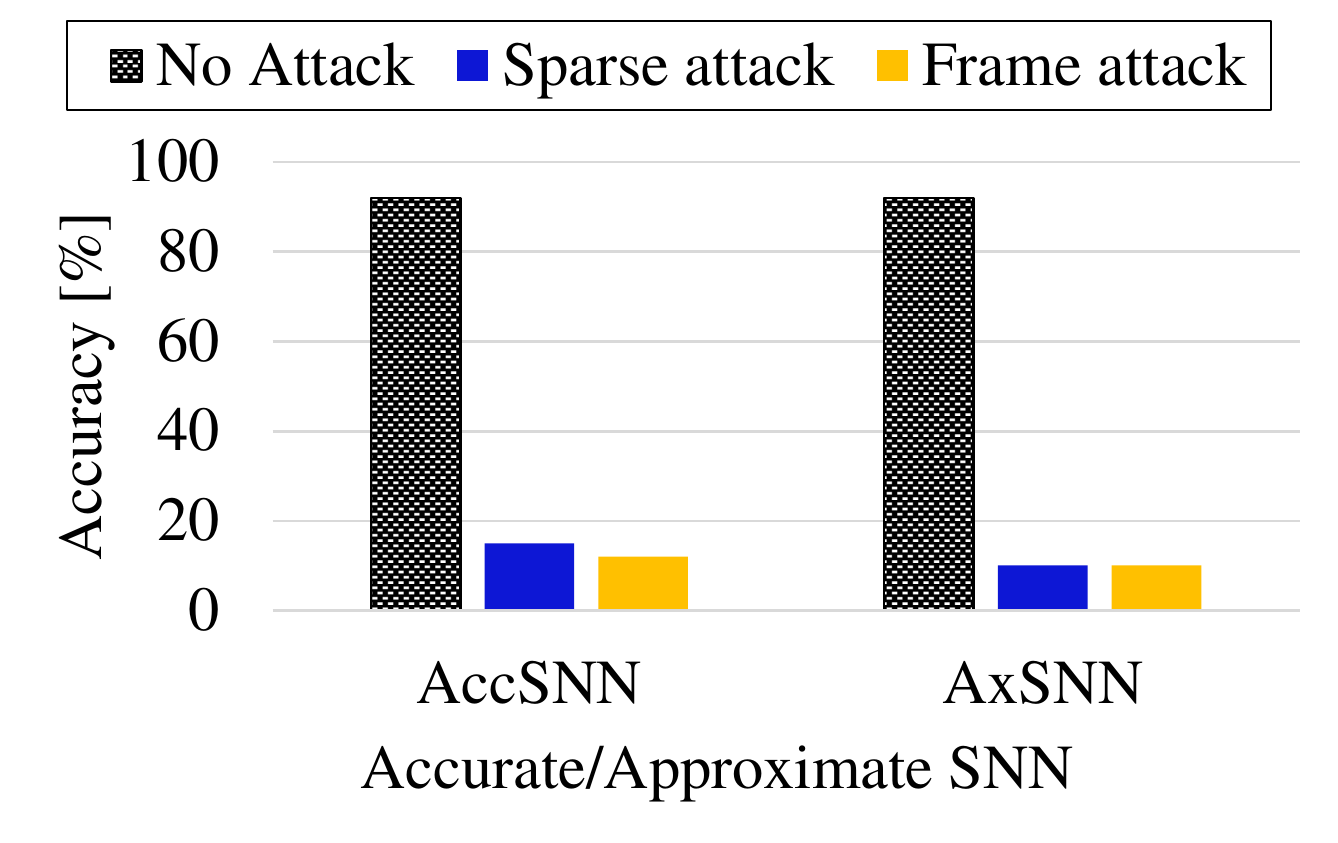}
		\label{fig:neuromorphic}
	\end{subfigure}
    \vspace{-0.4in}
    \caption{Accuracy of AccSNN without attack ($\epsilon$=0) for MNIST and accuracy of AccSNN and AxSNN for DVS128 Gesture with and without attacks.} 
    \label{fig:noattackandgraph}
    \vspace{-0.3in}
\end{figure}
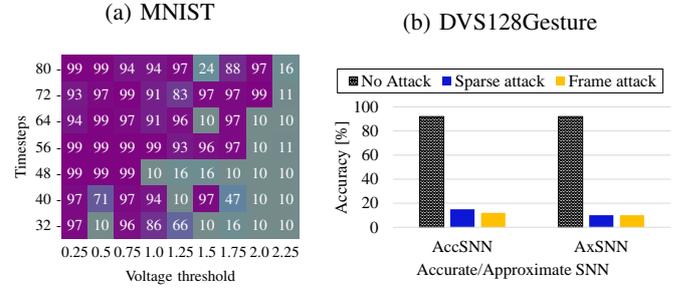

For evaluating the precision-scaling-based adversarial defense in AxSNNs for the static MNIST dataset, we start with the approximation level of 0.01 and vary the precision-scaling scales as Int8, FP16, and FP32. Specifically, we analyze the robustness of each precision-scaled AxSNN against PGD and BIM attacks, with perturbation budget ($\epsilon$=1.0), by comparing the accuracy corresponding to each precision scale in Fig. \ref{fig:fp32}, Fig. \ref{fig:fp16}, and Fig. \ref{fig:int8} with the base model accuracy (AccSNN) in Fig. \ref{fig:noattack} for the MNIST classification. We used $\epsilon$=1.0 because the accuracy of both AccSNN and AxSNN drops significantly after this value, so it becomes non-recoverable. For example, it drops to 10\% with $\epsilon$=1.5. 

Furthermore, we observe that the robustness of precision-scaled AxSNNs varies while we change parameters such as threshold voltage, time steps, precision-scaling scale, and approximation level. For example, a PGD attack ($\epsilon = 1.0$) results in a 12\% accuracy loss in precision-scaled AxSNN with approximation level 0.01, precision scale FP32, threshold voltage 0.75 and time steps 32, when compared to AccSNN (see Fig. \ref{fig:noattack}). However, changing the precision scale to FP16 and INT8 results in 7\% (see Fig. \ref{fig:pgdfp16}) and 4\% accuracy loss (see Fig. \ref{fig:pgdint8}, respectively, in the same AxSNNs when compared to AccSNN (see Fig. \ref{fig:noattack}). Likewise, changing threshold voltage and time steps to 1.0 and 48, respectively, results in recovering the accuracy of AxSNNs to 95\%. Similarly, a BIM attack ($\epsilon = 1.0$) results in a 15\% accuracy loss in precision-scaled AxSNN with approximation level 0.01, precision-scaled FP32, threshold voltage 0.5 and time-step 72, when compared to AccSNN (see Fig. \ref{fig:noattack}). However, changing the precision scale to FP16 and INT8 results in 4\% (see Fig. \ref{fig:pgdfp16}) and 3\% accuracy loss (see Fig. \ref{fig:pgdint8}, respectively, in the same AxSNN when compared to the AccSNN (see Fig. \ref{fig:noattack}). Similarly, changing threshold voltage and time steps to 1.25 and 64, respectively, results in recovering the accuracy of AxSNNs to 97\%. This indeed highlights the importance of using our proposed Algorithm \ref{alg:robust} for finding a robust sweet spot for the most suitable threshold voltage, time steps, approximation level, and precision scale under an attack. It is important to note that some deviating behavior from the above-discussed robustness trend is often observed at very high threshold voltage, i.e., greater than 1.75, and high approximation level, i.e., greater than 0.1. It is not surprising because SNNs typically lack performance efficiency for very high threshold voltage and high approximation levels since the spikes that trigger the LIF neurons may not cross the threshold voltage, resulting in a compromised performance of the whole network.

We test our proposed Algorithm \ref{alg:robust} using threshold voltage in the range of 0.25 to 2.25 with an interval of 0.25, time steps in the range of 32 to 80 with an interval of 8, and precision scales as Int8, FP16, and FP32. As shown in Table \ref{tab:bestlevel}, our Algorithm \ref{alg:robust} identifies the best parameter configurations even under PGD and BIM attacks. For example, our algorithm determines the approximation level 0.01 for precision-scaled AxSNN with the precision scale FP32, threshold voltage 1.0, and time steps 48 for accuracy of 97\% even under PGD attack ($\epsilon = 1.0$). However, it determines an approximation level 0.011 for precision-scaled AxSNN with the precision scale INT8, threshold voltage 0.75, and time steps 32 for accuracy of 88\% under the same attack. It is worth mentioning that the robustness of AxSNNs varies with the precision scales, and thus our algorithm provides different approximation levels for a combination of different threshold voltage, time steps, and precision scales. For instance, a PGD attack ($\epsilon = 1.0$) on AxSNN with approximation level 0.1, time steps 32, and threshold voltage 0.25 results in a 72\% accuracy loss, whereas the same attack on the precision-scaled AxSNN with approximation level 0.01, precision scale Int8 leads to only 17\% accuracy loss in the static MNIST dataset (4X robustness improvement). Similar trend is observed with the BIM attack.

\subsubsection{\uline{AQF-based Adversarial Defense}} 

\begin{table}[!t]
\centering
\caption{Best robustness settings for the proposed precision-scaled AxSNN classifier for the MNIST dataset.}
\scalebox{0.9}
{
\begin{tabular}{|l|l|l|l|}
\hline \rowcolor[HTML]{EFEFEF}
$(V_{th},T)$                 & Attacks & ($q_{th}$, $a_{th}$) & Accuracy[\%] \\ \hline
\multirow{2}{*}{(0.25,32)} & PGD     & (FP32, 0.01)        & 88       \\ \cline{2-4} 
                           & BIM     & (INT8, 0.009)      & 80       \\ \hline
\multirow{2}{*}{(0.75,32)} & PGD     & (INT8, 0.011)      & 92       \\ \cline{2-4} 
                           & BIM     & (FP16, 0.013)       & 91       \\ \hline
\multirow{2}{*}{(1.0,48)}  & PGD     & (FP32, 0.01)      & \textbf{97}       \\ \cline{2-4} 
                           & BIM     & (INT8, 0.0125)     & \textbf{96}       \\ \hline
\end{tabular}}
\label{tab:bestlevel}
\vspace{-0.2in}
\end{table}

To evaluate the effectiveness of the proposed AQF-based adversarial defense in the precision-scaled AxSNN, we use Algorithm \ref{alg:aqaf} for DVS128 Gesture classification under the sparse and frame attacks. Unfortunately, training AxSNNs takes a very long time; thus, we limit ourselves in testing the proposed Algorithm \ref{alg:aqaf} with threshold voltage 1.0 and time steps 80 only. Note, these parameter settings are the most common in SNN research community for neuromorphic datasets \cite{frady2020neuromorphic}. Table \ref{tab:bestlevel2} enlists the approximation levels identified by our algorithm for precision scales 0.015 and 0.01. We observe that a sparse attack on AxSNN leads to a 77\% accuracy loss when compared to AccSNN without any attack. However, AQF-based adversarial defense in precision-scaled AxSNN recovers the accuracy close to the baseline accuracy. For instance, incorporating AQF-based adversarial defense with a precision scale of 0.015 and approximation level of 0.1 in a precision-scaled AxSNN under a sparse attack recovers the accuracy to almost 90\%, which is only 2\% less than the baseline accuracy.
Interestingly, a frame attack on AxSNN leads to 82\% accuracy loss when compared to AccSNN. However, AQF-based adversarial defense with a precision scale of 0.015 and an approximation level 0.1 in a precision-scaled AxSNN under a frame attack recovers the accuracy to almost 91\%, which is only 1\% less than the baseline accuracy. The reason behind such a tremendous improvement in robustness is that AQF masks noisy events that have a low correlation with each other in a perturbed neuromorphic dataset.

%





\begin{table}[!t]
\caption{Recovered accuracy $A_r$ and accuracy loss $A_l$ when compared to baseline accuracy with AQF filtering in AxSNN ($V_{th}$, $T$) = (1.0, 80) using DVS128 Gesture.}
\centering
\scalebox{0.9}
{
\begin{tabular}{|l|l|l|l|}
\hline \rowcolor[HTML]{EFEFEF}
Attacks                        & ($q_t$, $a_{th}$) & $A_r${[}\%{]} &  $A_l${[}\%{]} \\ \hline
\multirow{4}{*}{Sparse Attack}  & (0.015, 0.1) & \textbf{90.01}& \textbf{2.0} \\ \cline{2-4} 
                               & (0.01, 0.15) & 88.4 &  3.6          \\ \cline{2-4} 
                               & (0.0, 0.001) & 84.3  &  7.7     \\ \hline
\multirow{4}{*}{Frame Attack}  & (0.015, 0.1) & \textbf{91.1} & \textbf{1.0}           \\ \cline{2-4} 
                                & (0.01, 0.15) & 89.9   &  2.1        \\ \cline{2-4} 
                                & (0.0, 0.001) & 88.2  &   3.8        \\ \hline
\end{tabular}}
\label{tab:bestlevel2}
\vspace{-0.25in}
\end{table}

\vspace{-3mm}
\section{Conclusion}
\label{sec:conclusion}
\vspace{-1mm}
In this paper, we extensively explored the adversarial robustness of AxSNNs and proposed two novel defense methods: precision scaling and approximate quantization-aware filtering (AQF) for designing adversarially robust AxSNNs. To demonstrate the effectiveness of these defense methods, we employ the static MNIST and the neuromorphic DVS128 Gesture datasets. Our results show that AxSNNs are more prone to adversarial attacks than AccSNNs, but precision scaling and AQF significantly improve the robustness of AxSNNs. For instance, a PGD attack on AxSNN results in a 72\% accuracy loss compared to AccSNN without any attack, whereas the same attack on the precision-scaled AxSNN leads to only a 17\% accuracy loss in the static MNIST dataset (4X robustness improvement). Likewise, a Sparse Attack on AxSNN leads to a 77\% accuracy loss when compared to AccSNN without any attack, whereas the same attack on an AxSNN with AQF leads to only a 2\% accuracy loss in the neuromorphic DVS128 Gesture dataset (38X robustness improvement).

\vspace{-1mm}
\section{ACKNOWLEDGEMENTS}
\vspace{-1mm}
This work was partially supported by awards from U.S. Naval Research Lab under the grant N0017321C2016. Any opinions, findings, and conclusions or recommendations expressed in this publication are those of the authors and do not necessarily reflect the views of the U.S. Government or agency thereof.
\vspace{-2mm}

\vspace{-2mm}
\bibliographystyle{IEEEtran}
\bibliography{bib/conf}

\end{document}